\documentclass[12pt,reqno]{article}
\usepackage{amsfonts}
\usepackage{amsmath}
\usepackage{amsbsy}
\usepackage{graphicx}

\usepackage{amssymb,latexsym}

 \numberwithin{equation}{section}

\begin{document}
 \allowdisplaybreaks[1]
\title{Constant Curvature Solutions of Minimal Massive Gravity}
\author{Nejat T. Y$\i$lmaz\\
Department of Electrical and Electronics Engineering,\\
Ya\c{s}ar University,\\
Sel\c{c}uk Ya\c{s}ar Kamp\"{u}s\"{u}\\
\"{U}niversite Caddesi, No:35-37,\\
A\u{g}a\c{c}l\i Yol, 35100,\\
Bornova, \.{I}zmir, Turkey.\\
\texttt{nejat.yilmaz@yasar.edu.tr}} \maketitle
\begin{abstract}
We study the constant curvature solutions of the minimal massive
gravity (MMGR). After introducing a condition on the physical and
the fiducial metrics as well as the St\"{u}ckelberg scalars which
truncates the action to the Einstein-Hilbert one in the presence
of an effective cosmological constant we focus on the solutions of
the constraint equations written for the constant curvature
physical metrics. We discuss two distinct formal solution methods
for these constraint equations then present an explicit class of
solutions for the St\"{u}ckelberg scalars and the fiducial metrics
giving rise to constant curvature solutions of MMGR.
\end{abstract}

\section{Introduction}
A Boulware-Deser (BD) \cite{BD1,BD2} ghost-free non-linear
extension of the Fierz-Pauli \cite{fp} massive gravity has been
constructed in a series of works in recent years. The entire
programme has originated from an approach which was constructed in
\cite{ah,crem} that enables the cancellation of the BD-ghost
order-by-order in the expansion of the mass term in terms of the
Goldstone bosons which is a consequence of the fact that the mass
term can be linked to the spontaneous symmetry breaking of the
general covariance. This method motivated de Rham, Gabadadze, and
Tolley to construct a two parameter family of massive non-linear
gravity actions free of ghost up to an order at the decoupling
limit \cite{dgrt1,dgrt2}. This construction was based on a fixed
flat background (fiducial or reference) metric which couples to
the physical metric in the graviton mass terms of massive
gravities via coordinate transformations. Later in
\cite{hr1,hr2,hr3} a generalized formulation of the dRGT theory
including a general background or fiducial metric has been
presented. In these later works it has also been shown that both
the dRGT as well as the so-called newly constructed minimal and
the general non-linear massive gravity (which was shown to contain
the dRGT action) theories are ghost-free at all orders.

In this work starting from the minimal massive gravity (MMGR)
action of \cite{hr1} written for the general background metric by
proposing a solution scheme which truncates the action to the
Einstein gravity on-shell we focus on the constant curvature
solutions of the theory. Thus we derive the constant curvature
Einsteinin sector in the MMGR. Following the introduction of a
constraint equation on the solutions which we show to be also
compatible with the field equations, as the total action is
truncated to the vacuum Einstein form with a cosmological constant
the constant curvature Einsteinian metrics become the solutions of
the MMGR. In this context instead of solving the coupled
St\"{u}ckelberg scalars, the constant curvature Einstein metric
and the fiducial metric directly from the highly non-linear field
equations one has to consider solving a matrix constraint equation
for any choice of constant curvature Einstein metric. After
formulating the integral form of these constraint equations which
boil down to be just scaled coordinate transformations between the
background fiducial and the physical metrics we will inspect two
distinct solution methodologies for finding non-trivial coordinate
transformations linking a set of fiducial metrics admitting a
special form and the constant curvature Einstein metrics. Finally
we will present a class of exact solutions of these coordinate
transformations namely the St\"{u}ckelberg scalars and the
corresponding fiducial metrics which eventually satisfy the field
equations when they couple to any constant curvature Einstein
metrics.

Section two is devoted to the construction of the solution scheme
of the MMGR field equations for constant curvature physical
metrics. In Section three we will obtain the integral form of the
constraint equations which become coordinate transformations on
the constant curvature Einstein metrics for diagonal form of
fiducial metrics. Section four contains the discussion about the
two distinct solution methodologies of these constraint equations.
Finally in Section five by using one of these methods we will
exactly integrate out the constraint equations for an explicitly
chosen form of the fiducial metric so that the resulting class of
solutions of St\"{u}ckelberg scalars and the fiducial reference
metric lead to the constant curvature solutions of the MMGR.
\section{The set-up}
 The minimal non-linear modification to the massive
Fierz-Pauli action \cite{fp} is constructed in \cite{hr1} which is
the minimal ghost free massive gravity action. It reads
\footnote{Here we slightly change notation and consider a global
form of the action rather than a coordinate basis one.}
\begin{equation}\label{e1}
 S_{MMGR}=-M_p^2\int\bigg[ R\ast 1+2m^2tr(\sqrt{\Sigma})\ast 1+\Lambda^{\prime}\ast
 1\bigg],
\end{equation}
 where $M_{p}$ is the planck mass, $m$ is the graviton mass, $R$ is the Ricci scalar, and the Hodge star operator is defined as
\begin{equation}\label{e2}
\ast 1=\sqrt{-g} dx^4,
\end{equation}
with $g=det(g_{\mu\nu})$. The effective cosmological constant is
\begin{equation}\label{e3}
\Lambda^{\prime}=\Lambda-6m^2.
\end{equation}
 In the
 above action we have defined the four by four matrix $\Sigma$ as
\begin{equation}\label{e4}
(\Sigma)^{\mu}_{\nu}=g^{\mu\rho}f_{\rho\nu},
\end{equation}
with $g^{\mu\nu}$ being the inverse metric and the four by four
matrix functional $f$ is
\begin{equation}\label{e5}
f_{\rho\nu}=\partial_{\rho}\phi^a\partial_{\nu}\phi^{b}\bar{f}_{ab}(\phi^{c}).
\end{equation}
Here\footnote{For the entire formulation below we will assume that
$\mu,\nu\cdots=\texttt{0,1,2,3}$ as well as
$a,b,c\cdots=\texttt{0,1,2,3}$ where as
$\textit{i,j,k}\cdots=\texttt{1,2,3}$.} $\phi^{a}$ with
$a,b,c=0,1,2,3$ are the generalized St\"{u}ckelberg scalar fields
which correspond to the coordinate transformations \cite{ah} and
$\bar{f}_{ab}(\phi^{c})$ is an auxiliary metric to be determined
(or chosen) depending on the physical problem at hand and it is a
function of the scalars. Following the original definition in
\cite{hr1} we have the square root matrix
\begin{equation}\label{e6}
\sqrt{\Sigma}\sqrt{\Sigma}=\Sigma.
\end{equation}
In the absence of matter the metric equation \cite{hr1} is
\begin{equation}\label{e7}
R_{\mu\nu}-\frac{1}{2}g_{\mu\nu}R-\frac{1}{2}\Lambda
g_{\mu\nu}+\frac{1}{2}m^2\big[g\sqrt{\Sigma}+\sqrt{\Sigma}^{\:T}g\big]_{\mu\nu}+m^2g_{\mu\nu}(3-tr\big[\sqrt{\Sigma}\:\big])=0.
\end{equation}
Also in \cite{hr1} the St\"{u}ckelberg scalar field equations are
equivalently formulated as
\begin{equation}\label{e8}
\nabla_{\mu}\big(\big[\sqrt{\Sigma}\:\big]^{\mu}_{\nu}+\big[g^{-1}\sqrt{\Sigma}^{\:T}g\big]^{\mu}_{\nu}-2tr\big[\sqrt{\Sigma}\:\big]\delta^{\mu}_{\nu}\big)
=0,
\end{equation}
where $\nabla_{\mu}$ is the covariant derivative of the
Levi-Civita connection acting on $\sqrt{\Sigma}$ as a (1,1)
tensor. Now consider solutions satisfying the constraint
\begin{equation}\label{e9}
\frac{1}{2}\big[g\sqrt{\Sigma}+\sqrt{\Sigma}^{\:T}g\big]-tr\big[\sqrt{\Sigma}\:\big]g=Cg,
\end{equation}
where $C$ is an arbitrary constant and we prefer the matrix
notation. When this solution constraint is substituted in
\eqref{e7} one obtains
\begin{equation}\label{e10}
R_{\mu\nu}-\frac{1}{2}g_{\mu\nu}R+\tilde{\Lambda} g_{\mu\nu}=0,
\end{equation}
where
\begin{equation}\label{e11}
\tilde{\Lambda}= -\frac{1}{2}\Lambda+3m^2+Cm^2.
\end{equation}
Therefore upon applying the constraint \eqref{e9} on the
St\"{u}ckelberg scalars, the physical and the fiducial metrics the
metric sector of MMGR is truncated to the cosmological constant
type vacuum-Einstein theory\footnote{At this stage there is no
guarantee whether such solutions are compatible with the field
equations and exist but below we will explicitly prove their
existence.}. Now \eqref{e9} implies
\begin{equation}\label{e12}
\sqrt{\Sigma}+g^{-1}\sqrt{\Sigma}^{\:T}g-2\,tr\big[\sqrt{\Sigma}\:\big]\mathbf{
1_{4}} =2C\mathbf{1_{4}},
\end{equation}
where $\mathbf{1_{4}}$ is the $4\times 4$ identity matrix. If we
take the covariant derivative of \eqref{e12} and contract it with
the upper indices of the matrices we see that solutions satisfying
this constraint automatically and trivially satisfy the
St\"{u}ckelberg scalar field equations \eqref{e8}. On the other
hand we also infer from the on-shell-redundant metric equation
\eqref{e10} that such solutions have the constant Ricci scalar
\begin{equation}\label{e13}
R=4\tilde{\Lambda}.
\end{equation}
So via \eqref{e10} we have the $\Lambda$-type Einstein metric
vacuum solutions whose Ricci curvature 2-forms are constant and
can be written as \cite{thring}
\begin{equation}\label{e14}
R^{\alpha\beta}= Ke^{\alpha\beta},
\end{equation}
for some moving co-frame ${e^{\alpha}}$. Here $R=12K$ and for our
solutions we have
\begin{equation}\label{e15}
K=\frac{1}{3}\tilde{\Lambda}.
\end{equation}
We observe that in this scheme one obtains the flat solutions if $
C=\Lambda/(2m^2)-3$. Now if we take the trace of \eqref{e12} we
find that
\begin{equation}\label{e16}
tr\big[\sqrt{\Sigma}\:\big]=-\frac{4}{3}C.
\end{equation}
Substituting this result back in \eqref{e12} we get the matrix
relation
\begin{equation}\label{e17}
g\sqrt{\Sigma}+\sqrt{\Sigma}^{\:T}g=-\frac{2C}{3}g.
\end{equation}
Using the symmetry of $g\sqrt{\Sigma}$ \cite{bac} and then by
taking the square of both sides we can more simplify this relation
to obtain
\begin{equation}\label{e17.5}
f=\frac{C^2}{9}\:g,
\end{equation}
which must be satisfied by any constant curvature metric solutions
of \eqref{e10}, the St\"{u}ckelberg scalar fields ${\phi^{c}}$ and
implicitly by the yet not specified fiducial metric $\bar{f}_{ab}$
so that they also satisfy \eqref{e7} and \eqref{e8} and become
least action solutions of the minimal massive gravity action
\eqref{e1}. \eqref{e17.5} is the most general form of the
constraint on such solutions to be solved. We observe that another
consequence of Eq. \eqref{e17.5} is the fact that the constant
curvature Einstein metric $g_{\mu\nu}$ which minimizes the action
\eqref{e1} must be a multiple of the induced metric which is the
pull-back of the fiducial metric $\bar{f}_{ab}(\phi^{c})$ via the
coordinate transformations generated by the St\"{u}ckelberg scalar
fields ${\phi^{c}}$. Next for a diagonal fiducial and constant
curvature physical metric parametrizations we will discuss
solution methods of this equation which can be considered as
fixing a gauge. This analysis will lead to functionally parametric
class of solutions for the St\"{u}ckelberg scalars and the
fiducial metric. Thus by finding non-trivial coordinate
transformations satisfying \eqref{e17.5} we will show that there
exists solutions of \eqref{e7} and \eqref{e8} which have constant
curvature that satisfy the pre-assumed constraint \eqref{e9}.
Explicit construction of a class of solutions in Section five will
justify and finalize the validity of our formalism suggested
above.
\section{St\"{u}ckelberg scalar equations}
Locally a space of constant curvature always has coordinates
${x^{\mu}}$ for which the corresponding metric can be written as
\cite{thring}
\begin{equation}\label{e18}
g_{\mu\nu}=\frac{\eta_{\mu\nu}}{\big(1+\frac{K}{4}x^2\big)^2},
\end{equation}
where $\eta_{\mu\nu}=diag(-1,1,1,1)$ and
$x^2=x^{\mu}x^{\nu}\eta_{\mu\nu}$. Now in component form
\eqref{e17.5} reads
\begin{equation}\label{e20}
\partial_{\mu}\phi^a\partial_{\nu}\phi^{b}\bar{f}_{ab}(\phi^{c})=\frac{C^2}{9}\:g_{\mu\nu}.
\end{equation}
We should state that in this equation only the diagonal components
are non-zero. To solve \eqref{e20} explicitly let us assume that
the fiducial metric is of the form
\begin{equation}\label{e21}
\bar{f}= \left(\begin{matrix}
f_{\texttt{00}}(\phi^{a})&\texttt{0}&\texttt{0}&\texttt{0}
\\
\texttt{0}&f_{\texttt{11}}(\phi^{a})&\texttt{0}&\texttt{0}\\\texttt{0}&\texttt{0}&f_{\texttt{22}}(\phi^{a})&\texttt{0}\\\texttt{0}&\texttt{0}
&\texttt{0}&f_{\texttt{33}}(\phi^{a})\end{matrix}\right).
\end{equation}
Following this choice and substituting \eqref{e18} in \eqref{e20}
we obtain the set of equations when $\mu=\nu$
\begin{equation}\label{e22}
\sum\limits_{a=\texttt{0}}^{\texttt{3}}\big(\partial_{\mu}\phi^a)^2f_{aa}(\phi^b)=C^{\prime}\frac{\eta_{\mu\mu}}{\big(1+\frac{K}{4}x^2\big)^2},
\end{equation}
and when $\mu\neq\nu$
\begin{equation}\label{e22.1}
\sum\limits_{a=\texttt{0}}^{\texttt{3}}\partial_{\mu}\phi^a\partial_{\nu}\phi^{a}f_{aa}(\phi^b)=0,
\end{equation}
where we define $C^{\prime}=C^2/9$. For the sake of generating
solutions in the  rest of our analysis instead of \eqref{e22.1} we
will consider the stronger conditions
\begin{equation}\label{e22.2}
\partial_{\mu}\phi^a\partial_{\nu}\phi^{a}=0,
\end{equation}
for each $a=\texttt{0,1,2,3}$ and again for $\mu\neq\nu$ by
bearing in mind that \eqref{e22.2} satisfy \eqref{e22.1}
trivially. Now by using \eqref{e22.2} one can show easily via
multiplying both sides of \eqref{e22} for each $\mu$ by
$\partial_{\alpha}\phi^{\alpha}\partial_{\beta}\phi^{\beta}\partial_{\gamma}\phi^{\gamma}$
where $\alpha,\beta,\gamma\neq\mu$ that \eqref{e22} leads for each
$a=\texttt{0,1,2,3}$ to the equations
\begin{equation}\label{e22.3}
\big(\partial_{a}\phi^a)^2f_{aa}(\phi^b)=C^{\prime}\frac{\eta_{aa}}{\big(1+\frac{K}{4}x^2\big)^2},
\end{equation}
where there is no sum on $a$. Inspecting \eqref{e22.3} one
realizes that $f_{\texttt{00}}$ must be negative. Keeping this
fact in mind, we have
\begin{subequations}\label{e23}
\begin{align}
\partial_{\texttt{0}}\phi^{\texttt{0}}\sqrt{-f_{\texttt{00}}} &=\pm
\frac{\sqrt{C^{\prime}}}{1+\frac{K}{4}x^2},\notag\\
\tag{\ref{e23}}\\
\partial_{\textit{i}}\phi^{\textit{i}}\sqrt{f_{\textit{ii}}} &=\pm
\frac{\sqrt{C^{\prime}}}{1+\frac{K}{4}x^2},\notag
\end{align}
\end{subequations}
where $\textit{i}=\texttt{1,2,3}$ are the spatial indices on which
there is no sum. These four equations must be solved together with
\eqref{e22.2} which define twenty four independent equations.
\section{Solution methodology}
One can consider two distinct methods to solve \eqref{e22.2} and
\eqref{e23} for the St\"{u}ckelberg scalar fields ${\phi^{c}}$.
The first root one can follow depends on the observation that
\eqref{e23} enable one to choose any functional form for the
fiducial metric components $f_{aa}(\phi^b)$ of \eqref{e21} then do
separation of variables and perform partial integration on both
sides. This leads to the set of equations
\begin{subequations}\label{e23.7}
\begin{align}
\int\sqrt{-f_{\texttt{00}}(\phi^{b})}\,\partial\phi^{\texttt{0}}
&=F^0_{\Delta},\notag\\
\tag{\ref{e23.7}}\\
\int\sqrt{f_{\textit{ii}}(\phi^{b})}\,\partial\phi^{\textit{i}}
&=F^{\textit{i}}_{\Delta},\notag
\end{align}
\end{subequations}
where $\Delta$ defines the region of spacetime and it is
$\Delta=C^{\prime\prime}C^{\prime\prime\prime}$ with the
definitions
\begin{subequations}\label{e25}
\begin{align}
\text{for $\mu=\texttt{0}$}:\quad\quad\quad\notag\\
C^{\prime\prime}&=-\frac{K}{4}\quad\quad,\quad\quad C^{\prime\prime\prime}=\frac{K}{4}x^ix_i+1,\notag\\
\tag{\ref{e25}}\\
\text{for $\mu=\textit{i}=\texttt{1,2,3}$}:\notag\\
C^{\prime\prime}&=\frac{K}{4}\quad\quad,\quad\quad
C^{\prime\prime\prime}=\frac{K}{4}\big(-(x^0)^2+\sum\limits_{j\neq
i}x^jx_j\big)+1.\notag
\end{align}
\end{subequations}
The region dependent $F^{\mu}_{\Delta}$ functions on the RHS of
\eqref{e23.7} result from the partial integration of the RHS of
\eqref{e23} with respect to $x^{\mu}$. When $\Delta>\texttt{0}$
\begin{equation}\label{e26}
F^{\mu}_{\Delta}=\pm
\frac{2\sqrt{C^\prime}}{\sqrt{4\Delta}}\arctan\big(\frac{2C^{\prime\prime}x^\mu}{\sqrt{4\Delta}}\big)+C_{\Delta}^\mu(x^{\nu\neq\mu}),
\end{equation}
when $\Delta<\texttt{0}$
\begin{equation}\label{e27}
F^{\mu}_{\Delta}=\pm
\frac{\sqrt{C^\prime}}{\sqrt{-4\Delta}}\ln\Bigg|\frac{2C^{\prime\prime}x^\mu-\sqrt{-4\Delta
}}{2C^{\prime\prime}x^\mu+\sqrt{-4\Delta}}\bigg|+C_{\Delta}^\mu(x^{\nu\neq\mu}),
\end{equation}
and when $\Delta=\texttt{0}$
\begin{equation}\label{e27.1}
F^{\mu}_{\Delta}=\pm
\frac{\sqrt{C^\prime}}{C^{\prime\prime}x^\mu}+C_{\Delta}^\mu(x^{\nu\neq\mu}),
\end{equation}
where $C_{\Delta}^\mu(x^{\nu\neq\mu})$ are integration
variable-independent functions of the partial integrations. The
method one has to follow in finding solutions to \eqref{e23.7}
subject to the conditions \eqref{e22.2} is first to choose
functional forms for $f_{aa}(\phi^b)$ for each
$a=\texttt{0,1,2,3}$, perform the partial integration on the LHS
of \eqref{e23.7} and solve algebraically $\phi^a$'s. Then one can
apply the constraints \eqref{e22.2} to solve
$C_{\Delta}^\mu(x^{\nu\neq\mu})$'s. We must remark a couple of
important points here. Firstly this analysis is done region by
region depending on $\Delta$. Secondly not all functional forms of
$f_{aa}(\phi^b)$ may lead solutions. One may have to inspect the
conditions \eqref{e22.2} carefully in order to choose sensible
functional dependence of the fiducial metric components on the
St\"{u}ckelberg scalars so that there exist solutions. In this
direction for example one may assume that $f_{aa}=f_{aa}(\phi^a)$
only. Finally one should observe that the conditions for each
$\mu=\texttt{0,1,2,3}$
\begin{equation}\label{e27.2}
\partial_{\nu\neq\mu}\phi^{\mu}=0,
\end{equation}
solve \eqref{e22.2} and reduce the number of conditions to be
satisfied by each $\phi^{\mu}$ from six to three. Therefore they
are more preferable to construct the right form of the fiducial
metric components out of the St\"{u}ckelberg fields that will lead
to solutions. As an alternate to the above-mentioned method which
is based on the explicit choice of the functional dependence of
the fiducial metric components on the St\"{u}ckelberg scalars one
can consider their implicit dependence as well namely one can
propose how $f_{aa}$'s depend on the spacetime coordinates
directly suppressing the St\"{u}ckelberg scalars which are just
coordinate transformations to an implicit level. In this direction
first for each $\mu=\texttt{0,1,2,3}$ let us introduce three
functions $G^{\mu}_j(x^{\nu})$ for $j=\texttt{1,2,3}$. Next we
will define the fiducial metric components as any functionals of
these functions
\begin{equation}\label{e27.3}
f_{\mu\mu}=f_{\mu\mu}[G^{\mu}_{\texttt{1}}(x^{\nu}),G^{\mu}_{\texttt{2}}(x^{\nu}),G^{\mu}_{\texttt{3}}(x^{\nu})].
\end{equation}
When one specifies the form of these functionals in terms of the
yet undetermined functions and substitute them in \eqref{e23} one
can do the partial integration which yields the St\"{u}ckelberg
scalars as
\begin{subequations}\label{e27.4}
\begin{align}
\phi^{\texttt{0}}(x^{\nu}) &=\pm
\int\frac{\sqrt{C^{\prime}}\,\partial x^{\texttt{0}}}{\sqrt{-f_{\texttt{00}}[G^{\texttt{0}}_{j}(x^{\nu})]}\big(1+\frac{K}{4}x^2\big)},\notag\\
\tag{\ref{e27.4}}\\
\phi^{\texttt{\textit{i}}}(x^{\nu}) &=\pm
\int\frac{\sqrt{C^{\prime}}\,\partial
x^{\textit{i}}}{\sqrt{f_{\textit{ii}}[G^{\textit{i}}_{j}(x^{\nu})]}\big(1+\frac{K}{4}x^2\big)}.\notag
\end{align}
\end{subequations}
We should not forget that these coordinate transformations are
subject to the conditions \eqref{e22.2} which we have discussed to
be satisfied by the special choice \eqref{e27.2}. Therefore the
unknown functions $G^{\mu}_j(x^{\nu})$ can be determined by
applying \eqref{e27.2} for each St\"{u}ckelberg scalar field in
\eqref{e27.4}. If one fixes the resultant integration
variable-independent functions of the partial integration to be
zero applying \eqref{e27.2} in \eqref{e27.4} yields
\begin{subequations}\label{e27.5}
\begin{align}
\partial_{\textit{i}}\bigg(\frac{1}{\sqrt{-f_{\texttt{00}}[G^{\texttt{0}}_{j}(x^{\nu})]}\big(1+\frac{K}{4}x^2\big)}\bigg)=0,\notag\\
\tag{\ref{e27.5}}\\
\partial_{\nu\neq\textit{i}}\bigg(\frac{1}{\sqrt{f_{\textit{ii}}[G^{\textit{i}}_{j}(x^{\nu})]}\big(1+\frac{K}{4}x^2\big)}\bigg)=0.\notag
\end{align}
\end{subequations}
For each set of three functions
${G^{\mu}_1(x^{\nu}),G^{\mu}_2(x^{\nu}),G^{\mu}_3(x^{\nu})}$ for
$\mu=\texttt{0,1,2,3}$ one obtains three coupled non-linear first
order pde's. However we should state that these equations are
de-coupled in differentiation variable sense. Therefore as a
result when one chooses the functional forms in \eqref{e27.3} one
can substitute them in \eqref{e27.5}, then solve for
$G^{\mu}_j(x^{\nu})$'s from these pde's thus the fiducial metric
components are determined explicitly. Then one can substitute them
in \eqref{e27.4} and integrate out the St\"{u}ckelberg scalars.
\section{A class of solutions}
In this section we will present a class of solutions based on a
simple observation. The conditions in \eqref{e27.5} state that the
expressions inside the parenthesizes must depend on only the
spacetime variable of the corresponding St\"{u}ckelberg scalar
field index. In other words in \eqref{e27.5} the first parenthesis
must only depend on $x^{\texttt{0}}$ and the following ones only
on $x^{\texttt{1}},x^{\texttt{2}},x^{\texttt{3}}$ respectively.
Therefore if we set
\begin{equation}\label{e28}
f_{\mu\mu}=G^{\mu}_{\texttt{1}}G^{\mu}_{\texttt{2}}G^{\mu}_{\texttt{3}},
\end{equation}
then the following choice of functions
\begin{subequations}\label{e29}
\begin{align}
G^{\texttt{0}}_{\texttt{1}}&=\frac{1}{\big(1+\frac{K}{4}x^2\big)^2},\quad\quad
G^{\texttt{0}}_{\texttt{2}}=-1,\quad\quad
G^{\texttt{0}}_{\texttt{3}}=\frac{1}{\big(F_{\texttt{0}}(x^{\texttt{0}})\big)^2},\notag\\
\tag{\ref{e29}}\\
G^{\textit{i}}_{\texttt{1}}&=\frac{1}{\big(1+\frac{K}{4}x^2\big)^2},\quad\quad
G^{\textit{i}}_{\texttt{2}}=1,\quad\quad
G^{\textit{i}}_{\texttt{3}}=\frac{1}{\big(F_{\textit{i}}(x^{\textit{i}})\big)^2},\notag
\end{align}
\end{subequations}
where $F_{\mu}(x^{\mu})$ are arbitrary functions of a single
variable solve \eqref{e27.5}. In this case the fiducial metric
components become
\begin{equation}\label{e30}
f_{\texttt{00}}=\frac{-1}{\big(1+\frac{K}{4}x^2\big)^2\big(F_{\texttt{0}}(x^{\texttt{0}})\big)^2},\quad\quad
f_{\textit{ii}}=\frac{1}{\big(1+\frac{K}{4}x^2\big)^2\big(F_{\textit{i}}(x^{\textit{i}})\big)^2},
\end{equation}
and when these are substituted in \eqref{e27.4} we can solve the
St\"{u}ckelberg scalars as
\begin{equation}\label{e31}
\phi^{\texttt{0}}(x^{\texttt{0}}) =\pm
\int\sqrt{C^{\prime}}F_{\texttt{0}}(x^{\texttt{0}})dx^{\texttt{0}},\quad\quad
\phi^{\texttt{\textit{i}}}(x^{\textit{i}})=\pm
\int\sqrt{C^{\prime}}F_{\textit{i}}(x^{\textit{i}})dx^{\textit{i}}.
\end{equation}
If instead one chooses all $F_{\mu}(x^{\mu})=1$ then one obtains
the linear coordinate transformations
\begin{equation}\label{e32}
\phi^{\texttt{0}}(x^{\texttt{0}}) =\pm
\sqrt{C^{\prime}}x^{\texttt{0}}+C^{\texttt{0}},\quad\quad
\phi^{\texttt{\textit{i}}}(x^{\textit{i}})=\pm
\sqrt{C^{\prime}}x^{\textit{i}}+C^{\textit{i}},
\end{equation}
for which
\begin{equation}\label{e33}
f_{\texttt{00}}=\frac{-1}{\big(1+\frac{K}{4}x^2\big)^2},\quad\quad
f_{\textit{ii}}=\frac{1}{\big(1+\frac{K}{4}x^2\big)^2}.
\end{equation}
As a summary we have found that for any constant curvature
$\Lambda-$type vacuum Einstein solution metric which is
parameterized by $K$ via \eqref{e15} the St\"{u}ckelberg scalar
fields, and the fiducial metrics which form solutions of
\eqref{e23} and \eqref{e22.2} derived as a result of either method
discussed in Section four or as a special class of solutions
explicitly constructed in this section are also the constant
curvature solutions of the minimal massive gravity namely the
field equations \eqref{e7} and \eqref{e8}.
\section{Conclusion}
We have introduced a straightforward scheme to study the constant
curvature solutions of MMGR which replaces the method of directly
solving the metric and the St\"{u}ckelberg scalar field equations
upon fixing the fiducial metric by the method of solving a
constraint equation which is simplified to be a coordinate
transformation of the physical, and the fiducial (background)
metrics via the scalars of the mass term for a given constant
curvature Einstein metric. After obtaining the general form of
this constraint (coordinate transformation) equation on the
metrics which has no a-priori guarantee to be compatible with the
field equations we have simplified it by using the local form of
the constant curvature Einstein metrics which solve the on-shell
remainder part of the metric equation upon the substitution of our
constraint in the MMGR action. We then proceeded in discussing the
solution methods of finding non-trivial coordinate transformations
(gauges) which lead to the solutions of the theory by assuming a
diagonal form for the fiducial metric. Therefore under these
conditions we were able to construct an integral form of the
St\"{u}ckelberg scalars and thus the fiducial metric solutions
locally which can be integrated when one chooses the functional
form of the fiducial metric either in terms of the scalars or the
spacetime coordinates. We have also derived a class of explicit
solutions by integrating out these formal equations for the
St\"{u}ckelberg fields in the later method.

Though depending on a somewhat simple observation and approach we
should state that our construction enables one to find the
particular fiducial metric needed to solve the MMGR field
equations when one determines the desired physical metric
solutions of the MMGR. Moreover as it must be clear from Sections
three, four, and five one has the degree of freedom to arbitrarily
generate fiducial metrics which correspond to the same constant
curvature physical metrics by tuning the functional form of the
diagonal elements of the fiducial metric either in St\"{u}ckelberg
fields or spacetime dependence sense. This methodology is quite
different from the currently preferred one in the literature which
fixes the reference metric from the beginning and then studies the
solutions of the resulting MMGR action. This methodology may help
to cure the current challenge of finding stable thus physically
interesting solutions to the minimal and the general massive
gravity theories constructed in \cite{hr1} for a general fiducial
metric.

Studying Einstein sectors in massive gravities that is to say
inspecting how GR is embedded in the massive gravity theory which
is a modification to it can be enlightening in discovering and
understanding how the solutions of the modified theory perturbs
away from the ordinary GR solutions. This would serve to generate
predictions emerging from massive gravity to test its
validification in the form of corrections to the solutions of GR.
Therefore bearing in mind this physical intuition in this work we
have focussed on the constant curvature Einsteinian solutions of
the MMGR. The line of logic in our construction can be adopted to
shed light on other GR-type solutions as well. In particular one
may study the cosmological solutions for which the explicit form
of solutions derived in this work can be guiding. One may also
extend the solution method presented here to the more general form
of massive gravity to study its Einsteinian solutions.

\end{document}